\documentclass[aps,prl,reprint,preprintnumbers]{revtex4-2}
\usepackage{blindtext}
\usepackage{chngcntr}
\counterwithout{equation}{section} 
\usepackage{graphicx}
\usepackage{bbm}
\usepackage{amsmath}
\usepackage{amssymb}
\usepackage{mathtools}
\usepackage{amsthm}
\usepackage{hyperref}
\usepackage{mathrsfs}
\usepackage{marvosym}
\usepackage{dsfont}
\usepackage{xcolor}
\usepackage{braket}
\usepackage{cleveref}
\usepackage{comment}
\usepackage{float}
\usepackage{fancybox}
\usepackage[skins,theorems]{tcolorbox}
\tcbset{highlight math style={enhanced,
		colframe=black,colback=white,arc=0pt,boxrule=1pt}}

\newcommand{\beq}{\begin{equation}}  \newcommand{\eeq}{\end{equation}}
\newcommand{\bal}{\begin{aligned}}   \newcommand{\eal}{\end{aligned}}
\def\beqa{\begin{eqnarray}}
	\def\eeqa{\end{eqnarray}}

\newcommand{\dd}{\mathrm{d}}

\def\LSP{\Lambda_{\text{sp}}}

\def\simleq{\; \raise0.3ex\hbox{$<$\kern-0.75em
		\raise-1.1ex\hbox{$\sim$}}\; }
\def\simgeq{\; \raise0.3ex\hbox{$>$\kern-0.75em
		\raise-1.1ex\hbox{$\sim$}}\; }

\theoremstyle{remark}

\newtheoremstyle{named}{}{}{\itshape}{}{\bfseries}{.}{.5em}{#3}
\theoremstyle{named}

\newcommand{\para}[1]{\par\vspace{2mm}\noindent\textbf{#1}\,---\,}

\hypersetup{
	colorlinks = true,
	citecolor = {red},
	linkcolor = {blue},
	urlcolor = {blue},
}

\begin{document}
	\title{A Universal Pattern in Quantum Gravity at Infinite Distance}
	\author{Alberto Castellano}
	\email{alberto.castellano@csic.es}
	\author{Ignacio Ruiz}
	\email{ignacio.ruiz@uam.es}
	\affiliation{Instituto de F\'{i}sica Te\'{o}rica UAM-CSIC and Departamento de F\'{i}sica Te\'{o}rica, Universidad Aut\'{o}noma de Madrid, Cantoblanco, 28049 Madrid, Spain}
	\author{Irene Valenzuela}
	\email{irene.valenzuela@cern.ch}
	\affiliation{Instituto de F\'{i}sica Te\'{o}rica UAM-CSIC and Departamento de F\'{i}sica Te\'{o}rica, Universidad Aut\'{o}noma de Madrid, Cantoblanco, 28049 Madrid, Spain}
	\affiliation{CERN, Theoretical Physics Department, 1211 Meyrin, Switzerland}
	
	\begin{abstract}
		Quantum gravitational effects  become significant at a cut-off species scale that can be much lower than the Planck scale whenever we get a parametrically large number of fields becoming light. This is expected to occur at any perturbative limit of an effective field theory coupled to gravity, or equivalently, at any infinite distance limit in the field space of the quantum gravity completion. In this note, we present a universal pattern that links the asymptotic variation rates in field space of the quantum gravity cut-off $\LSP$ and the characteristic mass of the lightest tower of states $m$: $\frac{\vec\nabla m}{m}  \cdot\frac{\vec\nabla  \Lambda_{\rm sp}}{ \Lambda_{\rm sp}}=\frac1{d-2}$, where  $d$ is the spacetime dimension. This restriction can be used to make more precise several Swampland criteria that constrain the effective field theories that can be consistently coupled to quantum gravity.
		
	\end{abstract}
	\preprint{CERN-TH-2023-203,     IFT-UAM/CSIC-23-142}
	\maketitle

	\para{INTRODUCTION}	
	Effective Field Theories (EFTs) are very useful in High Energy Physics to describe the physical phenomena of our world. However, they are characterized by having a finite regime of validity, meaning that there is some cut-off energy scale at which the EFT breaks down and must be modified to incorporate e.g. new physical degrees of freedom. In this paper, we are interested in the quantum gravity cut-off scale at which an EFT weakly coupled to classical Einstein's gravity breaks down, and how this happens. In other words, what is the scale at which quantum gravitational effects become significant?
	
	The first naive guess is to set this scale to the Planck mass (of order $10^{19}$ GeV), since this determines the strength of the gravitational coupling. However, in certain cases, quantum gravitational effects can become significant at a cut-off scale which is much lower than the Planck scale. This occurs, for instance, when we have many light fields weakly coupled to gravity (termed \emph{species}), which renormalize the graviton propagator and lower the quantum gravity cut-off to $\LSP=M_{\rm Pl}/\sqrt{N}$ with $N$ being the number of species. Above this energy scale $\LSP$ (known as the species scale), quantum gravitational effects kick in and it is no longer possible to have a local EFT description weakly coupled to gravity. This scale is further motivated by black hole physics \cite{Dvali:2007wp, Dvali:2007hz}, unitarity of scattering amplitudes \cite{Donoghue:1994dn,Aydemir:2012nz,Anber:2011ut,Calmet:2017omb,Han:2004wt,Caron-Huot:2022ugt} and string theory \cite{vandeHeisteeg:2022btw,vandeHeisteeg:2023ubh,vandeHeisteeg:2023dlw,Castellano:2023aum}. 
	
	Clearly, the species scale can be made arbitrarily small $\LSP\ll M_{\rm Pl}$ whenever we get a parametrically large number of species. This is known to occur whenever we get an infinite tower of states becoming light (either because they open up some extra dimension, or because they correspond to oscillator modes of a weakly coupled higher dimensional object like a string). From a string theory perspective, the presence of a light tower of states is a universal feature that occurs whenever we try to engineer an exact global symmetry, since this pushes us to the boundaries of the field space. However, the existence of these towers acting as a censorship mechanism to restore global symmetries is expected to be a general feature of quantum gravity (even beyond string theory) and plays a central role in the Swampland program \cite{Brennan:2017rbf,Palti:2019pca,vanBeest:2021lhn,Grana:2021zvf,Harlow:2022gzl,Agmon:2022thq,VanRiet:2023pnx,Grimm:2018ohb, Gendler:2020dfp, Corvilain:2018lgw, Heidenreich_2021,Lanza:2020qmt,Lanza:2021udy}. The absence of exact global symmetries in quantum gravity has been shown using AdS/CFT \cite{Harlow:2018jwu,Harlow:2018tng,Harlow:2018fse} and black hole physics \cite{Banks:2010zn,Susskind:1995da,Israel:1967za}, in addition to the string theory evidence. Understanding in a quantitative way the behaviour of the tower of states would allow us to quantify how approximate a global symmetry can be and put sharp bounds on the value of gauge couplings and axionic decay constants (since the limit of a vanishing gauge coupling is equivalent to restoring a global symmetry). This can have important phenomenological implications for beyond Standard Models in Particle physics and Cosmology.   
	
	In this note we present a precise constraint that relates the asymptotic variation rates of the characteristic mass of the leading (i.e. lightest) tower of states, $m_{\rm t}$, and the species scale $\LSP$ as follows
	\beq \label{eq:patternmass}
	\frac{\vec\nabla m_{\text{t}}}{m_{\text{t}}} \cdot\frac{\vec\nabla \LSP}{\LSP}= \frac{1}{d-2}\, ,
	\eeq
	where $d$ is the spacetime dimension of our theory. 
	The variation rates are taken with respect to the space of couplings of the EFT as we take a perturbative limit in which some of these couplings become parametrically small. Since by definition, $m_{\rm t} \leq \LSP$, we will see that this pattern leads to sharp bounds on how fast the tower of states can become light and the value of the species scale at which quantum gravitational effects become significant. In the context of string theory, all coupling constants are controlled dynamically by the vacuum expectation values (vevs) of scalar fields, so the derivatives are taken with respect to those, as we explain below. Therefore, the pattern \eqref{eq:patternmass} can also constrain the scalar field variation that can be described by an EFT weakly coupled to gravity, yielding bounds of phenomenological interest for cosmological models of inflation or quintessence \cite{Scalisi:2018eaz}, as well as other dynamical proposals to explain the electro-weak hierarchy problem like cosmological relaxation \cite{Graham_2015}.

	In a companion paper \cite{pattern} we present strong evidence on its favour by analyzing a plethora of string theory constructions in different dimensions and with all allowed levels of supersymmetry. This pattern makes manifest an underlying structure behind all the different string theory examples that highly constrains the structure of the possible towers of states and helps to make more precise the Distance conjecture \cite{Ooguri:2006in,Etheredge:2022opl} in the Swampland program. In this paper, we introduce the pattern as well as some of its most immediate consequences, highlighting the key ingredients that make it work and providing  the first steps towards a bottom-up explanation for the constraint.

	\para{SYSTEMATICS OF THE PATTERN}	
	We consider in what follows a $d$-dimensional EFT containing some massless scalars (moduli), weakly coupled to gravity as follows
	\begin{equation}
		\mathcal{L}_{\text{EH + scalar}} = \dfrac{1}{2\kappa_d^2}\left(\mathcal{R} + \mathsf{G}_{i j} (\phi)\, \partial \phi^i \cdot \partial \phi^j\right)\, ,
	\end{equation}
	where $\mathcal{M}_{\phi}$ is the moduli space of the theory, namely the space of physically distinct vacua parametrized by the massless scalar field vevs $\braket{\phi^i}$. When probing any infinite distance boundary of $\mathcal{M}_{\phi}$, the Distance Conjecture \cite{Ooguri:2006in} requires from the appearance of at least one infinite tower of states becoming exponentially light. Therefore, in terms of the traversed \emph{geodesic} distance, which is defined by
	\begin{equation}\label{eq:modspacedist}
		\Delta_{\phi} = \int_{\gamma} \dd\sigma \sqrt{\mathsf{G}_{i j} (\phi) \frac{\dd \phi^i}{\dd \sigma} \frac{\dd \phi^j}{\dd \sigma}}\, ,
	\end{equation}
	with $\gamma$ denoting some geodesic path and $\sigma$ an affine parameter, there should exist a tower whose mass scale decreases as $m\sim e^{-\lambda \Delta_{\phi}}$ for $\Delta_{\phi} \gg 1$ (in Planck units), with $\lambda$ some $\mathcal{O}(1)$ factor. 
	
	In the presence of several moduli, it is useful to define a $\zeta$-vector for each tower becoming light, whose components read as
	\begin{equation}\label{eq:chargetomass}
		\zeta^i := - \mathsf{G}^{i j} \frac{\partial}{\partial \phi^j} \log m= -\partial^i \log m\, .
	\end{equation}
	These are denoted \emph{scalar charge-to-mass vectors} \cite{Calderon-Infante:2020dhm,Etheredge:2022opl,Etheredge:2023odp}, and they encode the information about how fast the associated tower of states becomes light. In particular, for any given asymptotically geodesic direction in moduli space characterized by some normalized tangent vector $\hat{T}$, the decay rate of the tower can be simply determined as the projection $\lambda=\vec{\zeta} \cdot \hat{T}$. Moreover, for any such limit, we will denote by $\vec{\zeta}_{\rm t}$ the scalar charge-to-mass vector of the leading (i.e. lightest) tower.
	
	Relatedly, the presence of an increasing number of light particle species in the theory implies a drastic breakdown of the EFT we started with. The maximum energy scale at which such description holds is known as the species scale $\LSP$ (see e.g. \cite{Castellano:2022bvr} and references therein), which in general is given by
	\beq
	\label{LSP}
	\LSP \simeq \frac{M_{\text{Pl};\, d}}{N^{\frac{1}{d-2}}}\, ,
	\eeq
	where $N$ is roughly the number of species at or below the species scale itself. Notice that whenever $N$ grows large at infinite distance, the species scale goes to zero (exponentially) in Planck units. To characterize how this occurs, one analogously introduces a $\mathcal{Z} $-vector as follows \cite{Calderon-Infante:2023ler}
	\begin{equation}\label{eq:speciescalechargetomass}
		\mathcal{Z}^i_{\text{sp}} := -\partial^i \log \LSP\, ,
	\end{equation}
	thus providing the asymptotic decay rate of the species scale. 
	
	In principle, there is some degree of independence between the vectors $\vec{\zeta}_{\text{t}}$ and $\vec{\mathcal{Z}}_{\text{sp}}$, meaning that one can get very different casuistics upon exploring different asymptotic corners of the moduli space. This is so since the species scale knows a priori about \emph{all} towers of (sufficiently) light states, such that the leading one does not always determine $\LSP$ alone. Nevertheless, the idea that we want to put forward in the present paper is that there seems to be a very precise link between these two quantities via the (asymptotic) relation 
	\begin{equation}\label{eq:pattern}
		\vec{\zeta}_{\rm t} \cdot \vec{\mathcal{Z}}_{\text{sp}} = \mathsf{G}^{ij} \left(\partial_i \log m_{\rm t}\right) \left(\partial_j \log \LSP\right)= \frac{1}{d-2}\; .
	\end{equation}
	This pattern holds universally regardless of the nature of the infinite distance limit that is explored as well as the microscopic interpretation of the towers. Using \eqref{LSP}, we can equivalently rewrite the above relation as follows
	\beq
	\label{patternN}
	\frac{\vec\nabla m_{\text{t}}}{m_{\text{t}}} \cdot\frac{\vec\nabla N}{N}=-1\, ,
	\eeq
	where the product is again taken with respect to the field space metric. Hence, the more fields we get, the slower they become light, in a very concrete way that is even independent of the spacetime dimension. 
	
	In the following, we will explain via some (realistic) toy model how this comes about, as well as commenting on the consequences that derive immediately from eq. \eqref{eq:pattern}. 
	In a companion paper \cite{pattern}, we will present string theory evidence in a wide range of scenarios supporting the claim. 
	
	\para{A simple Toy Model}
	Let us first show how the pattern works in simple single-modulus examples. We consider two cases: First, when dealing with a Kaluza-Klein decompactification of $n$ internal dimensions, we find a KK tower with characteristic mass $m_{\text{KK},\, n}$ yielding infinitely many states becoming light, with a spectrum of the form $m_k=k^{1/n} m_{\text{KK},\, n}$, where $k=1,\ldots,\infty$ \cite{Castellano:2021mmx}. Its associated species scale is the higher-dimensional Planck mass, which is given by $ M_{\text{Pl};\, d+n}=M_{\text{Pl};\, d}\, \left(\frac{m_{{\rm KK},\, n}}{M_{\text{Pl};\, d}}\right)^{\frac{n}{d+n-2}}$. Hence, the relevant charge-to-mass and species vectors, which can be obtained via dimensional reduction \footnote{The vectors in \eqref{eq:zeta&speciesveconemodulus} arise when considering decompactifications to a higher dimensional \emph{vacuum}. In the presence of warping, these are generically modified, see \cite{Etheredge:2023odp}.}, read as follows \cite{Calderon-Infante:2023ler}
	\beq\label{eq:zeta&speciesveconemodulus}
	\zeta_{{\rm KK},\, n} = \sqrt{\frac{d+n-2}{n (d-2)}}\, , \quad \mathcal{Z}_{{\rm KK}, \, n}=\sqrt{\frac{n}{(d+n-2) (d-2)}}\, .
	\eeq
	It is easy to check that these always reproduce the pattern \eqref{eq:pattern}, regardless of the number $n$ of decompactified dimensions. 
	
	Secondly, we can also get an infinite tower of states when having a higher dimensional object (like a weakly coupled string) becoming tensionless. In this case, the tower of string oscillator modes behaves as $m_k=\sqrt{k}\,m_{\rm osc}$ where $k=1,\ldots,\infty$, with an exponential degeneracy of states per level $k$. This leads to the following relevant vectors \cite{Etheredge:2023odp}
	\beq\label{eq:zeta&speciesvecstring}
	\zeta_{\rm osc}= \frac{1}{\sqrt{d-2}}=\mathcal{Z}_{\rm osc}\, ,
	\eeq
	since the species scale coincides with the string scale \cite{Castellano:2022bvr}. From this, it automatically follows that $\zeta_{\rm osc} \cdot \mathcal{Z}_{\rm osc} = \frac{1}{d-2}$, in agreement with \eqref{eq:pattern}. 
	
	However, this is not yet enough to show that the pattern holds in full generality, since when dealing with theories with more than one scalar field and more than one tower, the vectors $\vec{\zeta}_{\rm t}$ and $\vec{\mathcal{Z}}_{\rm sp}$ will not be necessarily parallel to each other. Still, the structure of the towers and the angles subtended by the vectors are always such that eq. \eqref{eq:pattern} is satisfied. For example, consider the case where several KK towers (associated to different internal cycles) become light. Then, the species scale is always given by some higher dimensional Planck mass, as in the one-modulus example above. For simplicity, we focus on a two-dimensional slice spanned by two KK towers decompactifying to $d+n$ and $d+n'$ dimensions, respectively, with associated canonically normalized volume moduli $\hat{\rho}$ and $\hat{\rho}'$. The $\zeta$-vectors are given by \cite{Etheredge:2022opl} 
	\begin{equation}\label{eq:n&n'zetas}
		\begin{split} 
			\vec{\zeta}_{{\rm KK},\, n} &= \left( 0 , \sqrt{\frac{d+n-2}{n (d-2)}} \right)\, ,\\
			\vec{\zeta}_{{\rm KK},\, n'} &= \left( \sqrt{\frac{d+n+n'-2}{n' (d+n-2)}} ,\, \sqrt{\frac{n}{(d+n-2)(d-2)}} \right)\, ,
		\end{split}
	\end{equation}
	whilst the relevant $\mathcal{Z}$-vectors are \cite{Calderon-Infante:2023ler}
	\begin{equation}
		\begin{split} 
			\vec{\mathcal{Z}}_{{\rm KK},\, n} &= \frac{n}{d+n-2} \vec{\zeta}_{{\rm KK},\, n}\, ,\\
			\vec{\mathcal{Z}}_{{\rm KK},\, n'} &= \frac{n'}{d+n'-2} \vec{\zeta}_{{\rm KK},\, n'}\, ,\\
			\vec{\mathcal{Z}}_{{\rm KK},\, n+ n'} &= \frac{ n \vec{\zeta}_{{\rm KK},\, n} + n' \vec{\zeta}_{{\rm KK},\, n'}}{d+n+n'-2} \, \label{eq:combinedZ}.
		\end{split}
	\end{equation}

	The species scale will correspond to the \emph{lightest} Planck scale for any chosen asymptotic trajectory $\hat T$ (i.e. that with the largest value of the exponential rate $\lambda_{\rm sp}=\vec{\mathcal{Z}}\cdot \hat T$). Hence, it will always be given by the Planck scale associated to full decompactification, $\vec{\mathcal{Z}}_{{\rm KK},\,  n+ n'}$, unless we move parallel to either one of the individual species vectors. The leading tower, on the other hand, will always be one of the two individual KK towers unless we move precisely parallel to $\vec{\mathcal{Z}}_{{\rm KK},\,  n+ n'}$, where all the internal geometry decompactifies in a homogeneous fashion. In any event, the pattern is always satisfied for any intermediate direction, since one can check that
	\beq\label{eq:doubleKK}
	\vec{\zeta}_{{\rm KK},\, n} \cdot \vec{\mathcal{Z}}_{{\rm KK},\, n+ n'}= \vec{\zeta}_{{\rm KK},\, n'} \cdot \vec{\mathcal{Z}}_{{\rm KK},\,n+ n'} = \frac1{d-2}\, .
	\eeq
	Similarly, when exploring some perturbative string limit in higher dimensions, the species scale will be given by the string scale, as in \eqref{eq:zeta&speciesvecstring} but the leading tower might be a KK tower rather than the tower of string oscillator modes. Upon restricting again to a 2d slice parametrized by the overall volume modulus and the $d$-dimensional dilaton, one finds the following relevant vectors (in a flat frame) \cite{Etheredge:2022opl,Calderon-Infante:2023ler}: 
	\begin{equation}\label{eq:zetasstringlimit}
		\begin{split}
			\vec{\zeta}_{\text{osc}}=\vec{\mathcal{Z}}_{\text{osc}} &= \left( \frac{1}{\sqrt{d+n-2}} , \sqrt{\frac{n}{(d+n-2)(d-2)}} \right)\, ,
		\end{split}
	\end{equation}
	whilst $\vec{\zeta}_{{\rm KK},\, n}$ and $\vec{\mathcal{Z}}_{{\rm KK},\, n}$ are computed as in \eqref{eq:combinedZ}. Recall that the leading tower (resp. species) becomes that with maximal projection along a given normalized tangent vector $\hat T$. Therefore, for intermediate directions (i.e. not aligned with any $\zeta$-vector in \eqref{eq:zetasstringlimit} above), $\LSP$ will be given by the string scale, while having the KK tower as the leading one. However, even in such case the pattern is still fulfilled, since

	\beq\label{eq:doubleKK2}
	\vec{\zeta}_{{\rm KK},\, n} \cdot \vec{\mathcal{Z}}_{\text{osc}}=\frac1{d-2}\, .
	\eeq
	Let us mention that all the previous considerations can be easily understood in a geometric way, upon depicting the different charge-to-mass and species vectors that enter in the game as illustrated below. Interestingly, despite the apparent simplicity of the previous ``toy models'' it turns out that all the different asymptotic corners of the moduli spaces arising from string theory constructions fit into one of these two sub-cases \cite{Lee:2019wij}. In fact, essentially the same type of pictures are always drawn, as analyzed in detail in the companion paper \cite{pattern}, differing only in how the diagrams are glued together in a way that respects the pattern, which puts non-trivial constraints on how different perturbative dual descriptions of the theory can fit together. This will also be further explored in \cite{taxonomyTBA}.

	\para{Derived bounds on the exponential rates}
	We would like to stress that a sharp relation like \eqref{eq:pattern} becomes rather constraining. In fact, several bounds in the Swampland literature immediately follow from imposing the pattern, as we now explain. First, eq. \eqref{eq:pattern} implies a \emph{lower} bound for the scalar charge-to-mass ratio of the leading tower of states. This is a direct consequence of the consistency condition $m_{\rm t}\leq \LSP$, from where one deduces that $|\vec{\zeta}_{\rm t} \cdot \vec{\mathcal{Z}}_{\text{sp}}| \leq |\vec{\zeta}_{\rm t}|^2$ (by Cauchy-Schwarz) and therefore
	\begin{equation}\label{eq:Rudelius}
		|\vec{\zeta}_{\rm t}|^2 \geq \frac{1}{d-2}\, .
	\end{equation}
	This leads precisely to the following lower bound for the exponential rate of the lightest tower
	\beq
	\lambda_{\text{t}} = |\vec{\zeta}_{\rm t}| \geq \frac{1}{\sqrt{d-2}}\, ,
	\eeq
	which was recently proposed in \cite{Etheredge:2022opl}. Relatedly, the fact that any infinite tower always satisfies the pattern with its own species scale implies, via the same argument, that there is an \emph{upper} bound for the decay rate of $\LSP$:
	\beq
	\lambda_{\text{sp}} = |\vec{\mathcal{Z}}_{\text{sp}}| \leq \frac{1}{\sqrt{d-2}}\, .
	\eeq
	This coincides with another recent proposal in \cite{vandeHeisteeg:2023ubh} based both in string theory examples and consistency of the EFT description \footnote{One can analogously obtain from \eqref{eq:pattern} the \emph{lower} bound for $\lambda_{\text{sp}} \geq \frac{1}{\sqrt{(d-1)(d-2)}}$ recently proposed in \cite{Calderon-Infante:2023ler} upon assuming that the maximum value for the exponential rate of the leading tower corresponds to that of a one-dimensional (unwarped) KK tower.}.
	
	Moreover, the pattern \eqref{eq:pattern} constrains the structure of the possible towers of states and how they can fit together as we move in moduli space. It is clearly related to the Emergent String Conjecture (ESC) \cite{Lee:2019wij}, which proposes that any infinite distance limit is either a decompactification or a perturbative string limit, since these are the obvious cases that fullfill the pattern. However, it is important that, when having several KK towers becoming light and signaling different decompactification limits, they can all be interpreted as KK towers in the \emph{same} dual frame. Otherwise, the pattern will not hold, as we further discuss later on.

	\begin{figure}[htb]
		\centering
		\includegraphics[scale=.50]{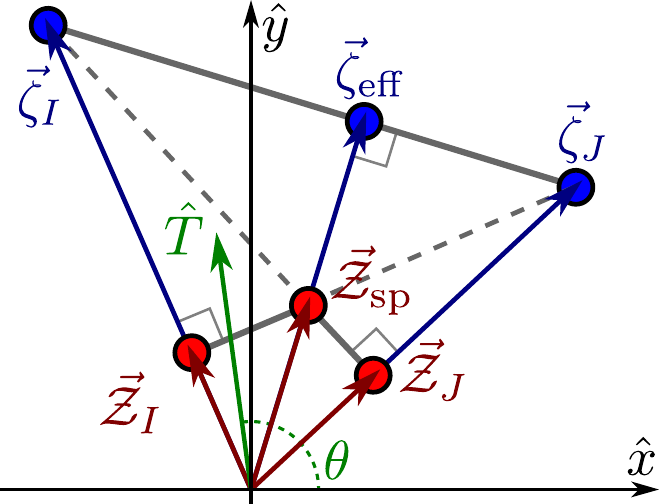}
		\caption{\small Sketch on how the pattern \eqref{eq:patternmass} and the $\zeta$-vectors associated to the leading towers $m_I$ and $m_J$, depending in some scalars $\hat{x}$ and $\hat{y}$, uniquely determine the \emph{multiplicative} species scale $\vec{\mathcal{Z}}_{\rm eff}$.} 
		\label{fig:ex}
	\end{figure}
	
	\para{Computing the quantum gravity cut-off} Let us now use the pattern to compute the quantum gravity cut-off upon knowledge of the behaviour of the mass of the tower but without assuming anything about its microscopic origin nor the associated density of states.
	
	Consider first a single tower of states with a scalar charge-to-mass vector \eqref{eq:chargetomass}, denoted by $\vec\zeta_I$. The species scale associated to this tower has $\vec{\mathcal{Z}}_I$ parallel to $\vec\zeta_I$, satisfying moreover \eqref{eq:pattern}. This scale sets the quantum gravity cut-off if we move along an asymptotic direction parallel to $\vec\zeta_I$, so that the exponential rates are related as 
	\beq
	\label{lsp}
	\lambda_{\rm sp}=\lambda_{\rm t}^{-1}/(d-2)\ .
	\eeq
	Notice that the structure and density of states of the tower determine the relation between $\LSP$ and $m_{\rm t}$, while the pattern \eqref{eq:pattern} further constraints that this relation is fully determined by the variation of the mass in the field space, so that  $\LSP\sim m_{\rm t}^{1/((d-2)\lambda_{\rm t}^2)}$ in this particular case.
	
	As we start moving along other asymptotic directions, there might be additional towers of states becoming light, thus contributing to the species scale. This is illustrated e.g. in Figure \ref{fig:ex}, where we have another tower with vector $\vec\zeta_J$. Hence, the species scale along any other intermediate asymptotic direction $\hat T$ will not be given by $\vec{\mathcal{Z}}_I$ but rather determined by another vector $\vec{\mathcal{Z}}_{\rm sp}$ that receives contributions from both towers. Interestingly, the pattern \eqref{eq:patternmass} determines \emph{uniquely} the species scale $\vec{\mathcal{Z}}_{\rm sp}$ once the mass of the leading towers is known.  First, notice that satisfying the pattern for both towers, i.e. $\vec{\zeta}_I\cdot \vec{\mathcal{Z}}_{\rm sp}=\vec{\zeta}_J\cdot \vec{\mathcal{Z}}_{\rm sp}=\frac{1}{d-2}$, implies that $\vec{\mathcal{Z}}_{\rm sp}$ must be perpendicular to $\vec{\zeta}_J-\vec{\zeta}_I$. 
	Secondly, the projection of  $ \vec{\mathcal{Z}}_{\rm sp}$ over $\vec{\zeta}_I$ must be precisely $\vec{\mathcal{Z}}_I$ since $\vec{\zeta}_I\cdot \vec{\mathcal{Z}}_{I}=\vec{\zeta}_I\cdot \vec{\mathcal{Z}}_{\rm sp}$. Finally, $ \vec{\mathcal{Z}}_{\rm sp}$ sets the value of the species scale as soon as we cease to move parallel to $\vec{\zeta}_I$, which is only consistent if the second tower $\vec{\zeta}_J$ starts contributing (i.e. its mass gets below $\Lambda_I$) at the same moment, implying that the projection of $\vec{\zeta}_J$ over $\vec{\zeta}_I$ must also be $\vec{\mathcal{Z}}_I$. All this determines the magnitude of $ \vec{\mathcal{Z}}_{\rm sp}$ and forces the vectors to be geometrically related as illustrated in Figure \ref{fig:ex}.  
	
	In summary, the exponential decay rate of the quantum gravity cut-off for any direction $\hat{T}=(\cos\theta,\sin\theta)$ within the cone spanned by $\vec{\zeta}_I$ and $\vec{\zeta}_J$ reads:
	\begin{equation}
		\lambda_{\rm sp}(\theta)=\hat{T}\cdot\vec{\mathcal{Z}}_{\rm sp}=\frac{1}{d-2}\frac{\left(\vec{\nabla}\log\frac{m_I}{m_J}\right)^\intercal\varepsilon\, \hat{T}}{\left(\vec{\nabla}\log m_I\right)^\intercal\varepsilon\, \vec{\nabla}\log m_J}\, ,
	\end{equation}		
	where for simplicity we work in a local basis of flat coordinates in the tangent bundle of the moduli space such that $\mathsf{G}_{ij}=\delta_{ij}$, and $\varepsilon=\bigl( \begin{smallmatrix}0 & -1\\ 1 & 0\end{smallmatrix}\bigr)$. This gets reduced to \eqref{lsp} in the particular case that $\hat T$ is parallel to $\vec \zeta_I$.
	
	Therefore, the maximum (geodesic) variation of the scalar fields that can be consistenly described by an EFT coupled to gravity in some perturbative corner reads:
	\beq
	\Delta\phi_{\text max}=\frac1{\lambda_{\rm sp}}\log{\frac{M_{\rm Pl}}{\LSP}}\, ,
	\eeq
	where we have used that the quantum gravity cut-off $\LSP$ decays exponentially with the field distance, and the exponential rate $\lambda_{\rm sp}$ can be either computed or bounded as explained above. If replacing $\LSP\leftrightarrow m_t$ and $\lambda_{\rm sp}\leftrightarrow \lambda_t$, we get the maximum scalar field range before we encounter the first state of the tower.
	
	\para{TOWARDS A BOTTOM-UP RATIONALE}	
	The pattern introduced in this note has been tested for a wide range of string theory compactifications, with different amounts of supersymmetry and diverse internal manifolds \cite{pattern}. It is natural to wonder whether this relation is a general feature of quantum gravity or just a lamppost effect of the string landscape. While we do not have yet a purely bottom-up argument (e.g. based on black hole arguments), we are still able to identify and motivate some sufficient conditions that allow the pattern to hold in a general way.
	
	The Distance Conjecture \cite{Ooguri:2006in} ensures that the mass of the leading tower -- and consequently, the species scale -- decreases \emph{exponentially} with the moduli space distance \eqref{eq:modspacedist}. This can be further motivated from a bottom-up perspective by the Emergence Proposal \cite{Palti:2019pca,vanBeest:2021lhn}, by which all the IR dynamics emerges from integrating out the dual massive degrees of freedom. This guarantees that $\vec{\zeta}_{\rm t}\cdot \vec{\mathcal{Z}}_{\rm sp}$ approaches a constant value asymptotically, but it does not constrain it to take the \emph{same} constant value $\frac{1}{d-2}$ for all infinite distance limits. To argue for this, we propose three \emph{sufficient} conditions which ensure that the pattern \eqref{eq:patternmass} is fulfilled along any asymptotic direction.

	\begin{center}
		\textbf{Condition 1}: \textit{The exponential rates $\lambda_I$ of the different towers $m_I$ are continuous over the asymptotic regions where they are defined. Furthermore, $\vec{\zeta}_{\rm t}\cdot\vec{\mathcal{Z}}_{\rm sp}$ must be well defined along any asymptotic direction}. 
	\end{center}

	This means that the exponential rate $\lambda_{\rm t}=\hat{T}\cdot\vec{\zeta}_{\rm t}$ of the leading tower is purely determined by the asymptotic direction $\hat{T}$, regardless of the particular geodesic we follow towards it. This does not imply that $\vec{\zeta}_{\rm t}$ has to remain constant along parallel trajectories, being allowed to change or \emph{slide} in the components perpendicular to $\hat{T}$ \footnote{Note that, given two parallel trajectories along which the tower becomes light with different exponential rates, as we move towards infinite distance $m_{\rm t}$ would take \emph{parametrically} distinct values between points separated a finite distance, rendering $\vec{\zeta}_{\rm t}$ with infinite norm. To avoid this, $\vec{\zeta}_{\rm t}$ should remain constant along parallel trajectories}. It implies, though, that the change in $\vec{\zeta}_{\rm t}$ has to be seen as a discrete \emph{jump} in terms of the asymptotic direction. This can occur either because: (1) the microscopic interpretation of the leading tower changes as a different tower starts dominating, in whose case the decay rate for both towers automatically coincides in the transition and $\lambda_{\rm t}$ is continuous, or (2) because a complicated moduli dependence of the mass makes $\vec{\zeta}_{\rm t}$ to \emph{jump} when crossing some \emph{sliding loci}  (see \cite{Etheredge:2023odp} for a detailed example when decompactifying to running solutions). In this latter case, we further need to require that $\vec{\zeta}_{\rm t}\cdot\vec{\mathcal{Z}}_{\rm sp}$ remains well-defined, otherwise the product will depend on the trajectory taken. The consequence of this condition is that we can divide the set of infinite distance limits into regions over which the $\vec{\zeta}_{\rm t}$ and $\vec{\mathcal{Z}}_{\rm sp}$ take fixed expressions, and thus their product is constant. 
	\begin{center}
		\textbf{Condition 2}: \textit{For every infinite distance limit along which several towers decay at the same rate, there must exist bound states involving all of them, such that the species scale must be given by the associated multiplicative species}. 
	\end{center}
	Consider several towers $\{m_1,\ldots,m_k\}$ becoming light at the same rate along some trajectory (or interface) with unit tangent vector $\hat{T}$, so that $\lambda_{\rm t}=\hat{T}\cdot\vec{\zeta}_{1}=\ldots=\hat{T}\cdot\vec{\zeta}_{k}$. These towers span a lattice of ``charges'' $(n_1,\ldots,n_k)$ given by the tower levels $m_{n_i,i}\sim n_i^{1/p_i}m_i$ (with $p_i>0$ depending on the nature and multiplicities of the tower \cite{Castellano:2021mmx}). If a (sub-)lattice of these charges is populated by particle states, then the total number of species is $N\sim \prod_{i=1}^k N_i$ (i.e. \emph{multiplicative}) rather than $N\sim \sum_{i=1}^k N_i$ (i.e. \emph{additive}). It can be shown \cite{Calderon-Infante:2023ler} that in the former case the resulting $\vec{\mathcal{Z}}_{\rm sp}$ is orthogonal to the hull spanned by the $\zeta$-vectors, and moreover dominates over the individual species scales. This implies that $\vec{\zeta}_1\cdot \vec{\mathcal{Z}}_{\rm sp}=\ldots= \vec{\zeta}_k\cdot \vec{\mathcal{Z}}_{\rm sp}$, such that the product \eqref{eq:pattern} takes the same value in the different adjacent regions (as well as in the interface). For additive species, though, we do not obtain any additional species vector, and $\vec{\zeta}_{\rm t}\cdot \vec{\mathcal{Z}}_{\rm sp}$ would generically change upon crossing the interfaces. This is why Condition 2 requires the existence of the (sub-)lattice of bound states yielding a multiplicative number of species, which can be further motivated by Swampland considerations \footnote{
		Notice that the scenario of additive species would result in independent towers of states becoming light at the same rate, naively implying different massless gravitons asymptotically, which goes against Swampland considerations \cite{Bedroya:2023tch,Kim:2019ths}. Furthermore, the case of multiplicative species is guaranteed by the Emergent String Conjecture \cite{Lee:2019wij} if assuming that all KK tower becoming light along some direction can be indeed microscopically interpreted as KK towers in the same dual frame.}.
	\begin{center}
		\textbf{Condition 3}: \textit{For every connected component of the space of infinite distance limits, there exists at least one direction associated to an emergent string limit or the homogeneous decompactification of an internal cycle to a higher dimensional vacuum}. 
	\end{center}
	With the previous two conditions, we have divided the moduli space into different regions and shown that  $\vec{\zeta}_{\rm t}\cdot\vec{\mathcal{Z}}_{\rm sp}$  remains constant across those. The only thing left is to set this constant to  $\frac{1}{d-2}$, which occurs if there exists \emph{at least one} asymptotic direction resulting in a string perturbative limit or a decompactification to a higher dimensional vacuum. This resembles but it is a weaker condition than the Emergent String Conjecture \cite{Lee:2019wij}.

	\para{CONCLUSIONS}	
	We propose a very concrete relation \eqref{eq:patternmass} between the quantum gravity cut-off $\Lambda_{\rm sp}$ in an EFT consistently coupled to quantum gravity, and the mass scale $m_{\rm t}$ of the lightest tower, which holds asymptotically in moduli space. At the moment, it is a common thread underlying all known string theory examples that have been explored so far. Finding a purely bottom-up rationale would have profound consequences for the consistency of EFTs coupled to gravity, since it constrains the possible towers of states predicted by the Swampland Distance conjecture and implies a precise lower bound on how fast they can become light. It also provides a clear recipe to determine the species quantum gravity cut-off upon knowledge of the leading tower of states, which puts further constraints on how different perturbative limits can fit together in the field space of a quantum gravity theory. \\

	The authors are supported by the Spanish Agencia Estatal de Investigacion through the grant ``IFT Centro de Excelencia Severo Ochoa'' CEX2020-001007-S and the grant PID2021-123017NB-I00, funded by MCIN/AEI/10.13039/ 501100011033 and by ERDF A way of making Europe. The work of A.C. is supported by the Spanish FPI grant No. PRE2019-089790 and by the Spanish Science and Innovation Ministry through a grant for postgraduate students in the Residencia de Estudiantes del CSIC. The work of I.R. is supported by the Spanish FPI grant No. PRE2020-094163. The work of I.V. is also partly supported by the grant RYC2019-028512-I from the MCI (Spain) and the ERC Starting Grant QGuide-101042568 - StG 2021.

	\bibliographystyle{apsrev4-1} 
	\bibliography{ref.bib}
	
\end{document}